\documentclass{aa} 
\usepackage{xcolor}
\usepackage{support-caption}
\usepackage{subcaption}
\usepackage{graphicx}
\usepackage[flushleft]{threeparttable}

\newcommand{\ha}{\ifmmode {\rm H}\alpha \else H$\alpha$\fi}
\newcommand{\hb}{\ifmmode {\rm H}\beta \else H$\beta$\fi}
\newcommand{\lya}{\ifmmode {\rm Ly}\alpha \else Ly$\alpha$\fi}
\newcommand{\pg}{\ifmmode {\rm P}\gamma \else Pa$\gamma$\fi}
\newcommand{\lyb}{\ifmmode {\rm Ly}\beta \else Ly$\beta$\fi}
\newcommand{\lyg}{\ifmmode {\rm Ly}\gamma \else Ly$\gamma$\fi}

\newcommand{\flyc}{\ifmmode \mathrm{f}_\mathrm{esc}\mathrm{(LyC)} \else $\mathrm{f}_\mathrm{esc}\mathrm{(LyC)}$\fi}

\def\kmsmpc{km s$^{-1}$ Mpc$^{-1}$}

\def\ergs{\ifmmode \mathrm{erg\hspace{1mm}s}^{-1} \else erg s$^{-1}$\fi}

\def\micron{\ifmmode \mu\mathrm{m} \else $\mu$m\fi}
\def\msun{\ifmmode \mathrm{M}_{\odot} \else M$_{\odot}$\fi}
\def\msunyr{\ifmmode \mathrm{M}_{\odot} \hspace{1mm}{\rm yr}^{-1} \else $\mathrm{M}_{\odot}$ yr$^{-1}$\fi}
\def\zsun{\ifmmode Z_{\odot} \else Z$_{\odot}$\fi}
\def\lsun{\ifmmode L_{\odot} \else L$_{\odot}$\fi}
\def\mstar{\ifmmode \mathrm{M}_{\star} \else M$_{\star}$\fi}

\usepackage{txfonts}
\newcommand{\orcid}[1]{\href{https://orcid.org/#1}{\includegraphics[width=10pt]{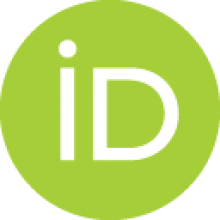}}}
\usepackage{hyperref}
\hypersetup{colorlinks, 
 linkcolor = [rgb]{1.,0.2,0.2},
 urlcolor = [rgb]{0.7,0.2,0.2},
 citecolor = [rgb]{0.1,0.4,1.},
 anchorcolor = [rgb]{0.7,0.2,0.2}
}
\begin{document} 


\title{Identifying \lya\ emitter candidates with Random Forest: learning from galaxies in CANDELS survey}

 \subtitle{}
 \author{L. Napolitano \orcid{0000-0002-8951-4408}
 \inst{1,2,3}
 \and L. Pentericci \orcid{0000-0001-8940-6768}
 \inst{1}
 \and A. Calabrò \orcid{0000-0003-2536-1614}
 \inst{1}
 \and P. Santini \orcid{0000-0002-9334-8705}
 \inst{1}
 \and M. Castellano \orcid{0000-0001-9875-8263}
 \inst{1}
 \and P. Cassata \orcid{0000-0002-6716-4400}
 \inst{5,6} \newline
 \and J. P. U. Fynbo \orcid{0000-0002-8149-8298}
 \inst{4}
 \and I. Jung \orcid{0000-0003-1187-4240}
 \inst{7}
 \and D. Kashino \orcid{0000-0001-9044-1747}
 \inst{9,10}
 \and S. Mascia \orcid{0000-0002-9572-7813}
 \inst{1,2,3} 
 \and M. Mignoli \orcid{0000-0002-9087-2835}
 \inst{8}
 }
 \institute{\textit{INAF – Osservatorio Astronomico di Roma, via Frascati 33, 00078, Monteporzio Catone, Italy}\\
 \email{lorenzo.napolitanoa@inaf.it}
 \and 
 \textit{Dipartimento di Fisica, Università di Roma Sapienza, Città Universitaria di Roma - Sapienza, Piazzale Aldo Moro, 2, 00185, Roma, Italy}
 \and 
 \textit{Dipartimento di Fisica, Università di Roma Tor Vergata, Via della Ricerca Scientifica, 1, 00133, Roma, Italy}
 \and
 \textit{The Cosmic Dawn Centre (DAWN), Niels Bohr Institute, University of Copenhagen, Lyngbyvej 2, DK-2100, Copenhagen, Denmark}
 \and
 \textit{Dipartimento di Fisica e Astronomia, Università di Padova, Vicolo dell’Osservatorio 3, I-35122, Padova, Italy}
 \and
 \textit{INAF – Osservatorio Astronomico di Padova, Vicolo dell’Osservatorio 5, I-35122, Padova, Italy}
 \and
 \textit{Space Telescope Science Institute, 3700 San Martin Drive Baltimore, MD 21218, USA}
 \and
 \textit{INAF – Osservatorio di Astrofisica e Scienza dello Spazio di Bologna, Via P. Gobetti 93/3, 40129 Bologna, Italy}
 \and
 \textit{Institute for Advanced Research, Nagoya University, Nagoya 464-8601, Japan}
 \and
 \textit{Department of Physics, Graduate School of Science, Nagoya University, Nagoya 464-8602, Japan}
}

 \date{Accepted XXX. Received YYY; in original form ZZZ}
 

\abstract{The physical processes which make a galaxy a Lyman Alpha Emitter have been extensively studied for the past 25 years. However, the correlations between physical and morphological properties of galaxies and the strength of the \lya\ emission line are still highly debated. Therefore, we investigate the correlations between the rest-frame \lya\ equivalent width and stellar mass, star formation rate, dust reddening, metallicity, age, half-light semi-major axis, Sérsic index and projected axis ratio in a sample of 1578 galaxies in the redshift range $2 \leq z \leq 7.9$ from the GOODS-S, UDS and COSMOS fields. From the large sample of \lya\ emitters (LAEs) in the dataset we find that LAEs are typically common main sequence star forming galaxies which show stellar mass $ \leq 10^9 \text{M}_{\odot}$, star formation rate $ \leq 10^{0.5} \text{M}_{\odot}/\text{yr}$, $E(B-V) \leq 0.2$ and half-light semi-major axis $\leq 1 \text{kpc}$. Building on these findings we develop a new method based on Random Forest (i.e. a Machine Learning classifier) in order to select galaxies which have the highest probability of being \lya\ emitters. When applied to a population in the redshift range $z \in [2.5, 4.5]$, our classifier holds a $(80 \pm 2)\%$ accuracy and $(73 \pm 4)\%$ precision. At higher redshifts ($z \in [4.5, 6]$) , we obtain a $73\%$ accuracy and a $80\%$ precision. These results highlight it is possible to overcome the current limitations in assembling large samples of LAEs by making informed predictions that can be used for planning future large scale spectroscopic surveys.}

 



 \maketitle

 
\section{Introduction} \label{sec:intro}
The \lya\ emission line is one of the brightest emission lines produced in star forming galaxies, due to the abundance of hydrogen and because it is produced by a common atomic electron transition.
At $z>2$ and $z>7$ the line shifts respectively to the optical and near-IR regime and it thus allows us to identity faint high-z objects very efficiently. Throughout 25 years of observations and discoveries \citep[e.g.,][]{Steidel1996, Stiavelli2001, Ouchi2003, Hayashino2004, Pentericci2018, Saxena2023}, \lya\ has shifted the observational redshift frontier, shedding light on the Epoch of Reionization (EoR) at $z>6$.\\
Traditionally \lya\ emitting galaxies are defined as \lya\ emitters (LAEs), if they show a rest-frame \lya\ equivalent width of $EW_0 \geq 20 \AA$ \citep[see for a reference][]{Shibuya2019, Ouchi_2020, Runnholm2020}. Searches for LAEs are often conducted through the use of narrow band (NB) filters \citep[with variable central $\lambda_{NB}$ and width $\delta_{NB}$ limited to $100 \AA$ - $200 \AA$, see][]{Cowie1998, Ajiki2003, Gronwall2007, Grove2009} that pinpoint the emission line in a certain redshift range targeted, i.e. $z = (\lambda_{NB} \pm \delta_{NB})/1216\AA -1$. In this case, the basic technique of finding LAE candidates involves comparing images taken through a narrow-band filter (which samples the flux from the emission line) with a broad-band one at close wavelengths (which samples the continuum emission). At increasingly high redshifts, the efficiency of this approach on ground based facilities is optimized by designing narrow band filters with wavelength centered in low background regions of the sky spectrum between the OH atmospheric emission lines, which begin to plague substantial wavelength ranges beyond $\lambda \sim 7000 \AA$. This highlights the main limitation of narrow band (NB) surveys: they probe very limited redshift ranges and hence, for a given survey area, small cosmological volumes. Moreover they can only uncover a fraction of the galaxy population which displays relatively bright \lya\ emission. In general a spectroscopic follow-up of a representative sample of the targets is mandatory to ensure the nature of the candidates. In fact NB surveys are subject to contamination of galaxies at lower redshifts that emit metal lines, such as CIV emission at $1549 \AA$ \citep[e.g.,][]{Fynbo2003}, MgII at $2798 \AA$ \citep[e.g.,][]{Dunlop2013_}, [OII] at $3727 \AA$ \citep[e.g.,][]{Fujita2003}, [OIII] at $5007 \AA$ \citep[e.g.,][]{Ciardullo2002}, which can fall in the same narrow band filter used to detect \lya\ from higher redshift sources. Other potential sources of contamination, which have to be considered if the narrow band and broad band images are taken in different periods, are transient objects, i.e. variable AGN or supernovae in the field \citep{Dunlop2013}.
Large samples of LAEs are also assembled trough spectroscopic observations which have the advantage of covering a large redshift range.
For example spectroscopy in a single observation extending over the wavelength range from $4000\AA$ to $8000\AA$ can unveil LAEs approximately from z=2.3 to z=5.6, thus allowing us to sample larger cosmological volumes. To date the deepest samples of LAEs come from this identification approach, successfully discovering LAEs reaching flux levels as low as a few $10^{-18}$ erg s$^{-1}$ cm$^{-2}$ \citep{Drake2017}. 
Because most of spectroscopic surveys \citep[e.g. VANDELS,][]{McLure2018, Pentericci2018, Garilli2021} are conducted though multi object spectrographs (MOS), which observe simultaneously only a relatively limited number of objects from few 10s to 100s in the best cases, to date target selection remains a key point. Only integral field unit spectrographs, such as MUSE, can observe   unbiased sample of galaxies' spectra, but  their small field of view limits their ability to work as wide survey probes. \\
The above considerations are at the bases of the present work, through which we try to understand  if it is possible to overcome the current limitations in assembling large samples of LAEs, making informed predictions on the bases of galaxies photometric and physical properties alone. The technique we propose is based on a Machine Learning (ML) algorithm which employ ensembles of decision trees, i.e. Random Forest classifier \citep{Breiman2001}. This approach relies on the fact that, as shown by many previous studies, the physical and morphological properties of LAEs are on average quite different from galaxies which do not show such a bright \lya\ emission (NLAEs) \citep[][]{Nakajima2012, Hagen2014, Ouchi_2020, McCarron2022}. In particular a key factor that shapes the final appearance of the \lya\ emission in a galaxy is dust which can absorb the UV continuum and \lya\ photons. The recurrent scattering nature of \lya\ \citep[for a review see][]{Dijkstra2017} driven by the neutral hydrogen gas actually increases the chance of the photon to be destroyed by dust absorption \citep{Verhamme2015, GurungLopez2022}.
As a result, even small amounts of dust can quench the \lya\ emission, resulting in the absence of the emission line. Thus the $N_{HI}$ column density of neutral hydrogen and the dust content 
are thought to be the most important physical quantities that determine the rate of escape of \lya\ photons. The major role of dust was demonstrated by previous observations which have reported that galaxies showing \lya\ in emission tend to have bluer UV continuum slopes ($\beta \sim -2$) than NLAEs \citep[][]{Shapley2001, Vanzella2009, Pentericci2009, Kornei2010}. 
This scenario also implies that the particular orientation of the emission path relative to the geometrical distribution of gas and dust in the emitting region should in principle be important to determine whether or not we are able to observe the line in emission. Theoretical models \citep[][]{Zheng2010, Verhamme2012, Behrens2014, Smith2019, Smith2022} predict a viewing angle effect where \lya\ photons escape more easily when the disk of the host galaxy is oriented face-on with respect to our line of sight. However clumpy structures were identified in high-z LAEs \citep[][]{Shibuya2016, Cornachione2018}, thus their morphological structure can not be modeled as a simple disk. This further complicates the viewing angle scenario of the \lya\ escape: subsequent studies are needed to explore this idea because the morphology of a galaxy is intrinsically linked to the \lya\ emission observed. 
Also the correlation between stellar mass and the presence of the \lya\ line is debated: \cite{Nakajima2012} reported that LAEs are likely to be low mass, faint continua galaxies; however results presented by \cite{Hagen2014} showed that LAEs are not exclusively low-mass sources. 
In order to get a clearer picture of the nature of LAEs, both physical and morphological properties have to be considered. 
In this context \cite{Paulino-Afonso2018} suggested a size evolution perspective: when the star formation is confined to a compact region $ \leq 1$ kpc, there are conditions to boost the escape of \lya\ photons to our line of sight, so that we observe the galaxy as a LAE. As time progresses, each galaxy grows in size, stellar mass, dust content, metallicity. Hence we measure less  \lya\  emission in larger galaxies, i.e. an  anti-correlation between \lya\ and the galaxy's size.\\
The physical processes that make a galaxy being a LAE are still highly debated, as well as the precise correlations between the galaxies' properties and the presence and strength of the emission line. In this work we therefore assemble a large sample of intermediate redshift galaxies with known spectroscopic, morphological and physical properties to further investigate the correlations between \lya\ emission and both physical (stellar mass, SFR, reddening, metallicity, age) and morphological (Sérsic index, half-light radius, projected semi-major axis) galaxies properties. We then use the same sample to train and test a new method to identify LAEs, based on supervised Machine Learning which builds on the existing correlations. The purpose is to select galaxies which have the highest probability of being LAEs from just the photometric information, which could for example drive informed target selection for future spectroscopic surveys.\\
To date, ML techniques have been successfully applied to remove contaminants from NB selected LAE candidates \citep{Ono2021} and to select LAEs in the HETDEX survey with an unsupervised learning approach \citep{Shanmugasundararaj2021}. An analysis of the physical properties (stellar mass, SFR and dust extinction) of 72 spectroscopically confirmed LAEs from the HETDEX survey was also carried out by \cite{McCarron2022} in order to predict the value of \lya\ EW for 10 LAEs at $z > 7$. Finally  \cite{Runnholm2020} used a linear regressor to predict \lya\ EW for 42 galaxies in the local Universe.
Compared to these works, our analysis will build on a much larger sample of galaxies whose \lya\ line was already measured through spectroscopy and we therefore aim to construct a robust method to identify LAEs from large surveys. \\
The paper is organized as follows. We describe the data set in Sec.~\ref{sec:data} and the methodology used in Sec.~\ref{sec:method}; we discuss the correlations found in our data in Sec.~\ref{sec:discussion}; in Sec.~\ref{sec:RandomForest} results of the Machine Learning method adopted are presented; we summarize our results and conclusions in Sec.~\ref{sec:conclusion}.
In the following, we adopt the $\Lambda$CDM concordance cosmological model ($H_0 = 70$ \kmsmpc, $\Omega_M = 0.3$, and $\Omega_{\Lambda} = 0.7$). 

\section{Data} \label{sec:data}
For our purposes we need to assemble the largest possible sample of sources which are associated with spectroscopic follow-up and with a measurement of both morphological and physical properties. In this sense the CANDELS survey \citep[][]{Grogin2011, Koekemoer2011} provides the optimal dataset: the 5 CANDELS field have homogeneous photometry obtained through HST observations (F125W, F160W and F814W in common), which is key for deriving unbiased morphological properties. These fields have been extensively studied thanks to many photometric and spectroscopic campaigns. Our sample  includes sources from only three of the CANDELS fields, namely GOODS-S, UDS and COSMOS, whose spectra are mostly publicly available. The use of three widely separated fields also have the advantage to mitigate cosmic variance yielding statistically robust samples of galaxies.

\subsection{Photometric catalogs and AGN removal}\label{sec:phot_cata}
For the UDS and COSMOS fields we used the  official photometric catalogs \citep[][respectively]{Galametz2013, Nayyeri2017} and photometric redshifts \citep{Kodra2023}, while in the case of GOODS-S, we adopted the updated 43-band catalog and photometric redshifts provided by \cite{Merlin2021}.
Since we are interested in searching for candidate \lya\ emitters amongst the star forming galaxy population, we flagged all the sources which show known X-ray emission, to remove AGN contaminants. For GOODS-S we used the AGN flags given by \cite{Luo2017}, for UDS the sources by \cite{Kocevski2018}  which have a $L_X > 10^{42}$ erg/s were removed \citep[see for example][]{Chen2017, Mukherjee2019}, while for COSMOS we relied on the galaxy classification flags given by each spectroscopic survey considered in Sec. ~\ref{sec:spec_cata}.

\subsection{Morphological catalogs}\label{sec:morph_cata}
We assumed that galaxies are well represented by a Sérsic profile \citep{sersicpaper}. 
For each galaxy we extracted the half-light semi-major axis ($R_e$), the Sérsic index ($n$) and the projected axis ratio ($q$) from the catalog by \cite{vanderWel2012} obtained by fitting the HST/WFC3 $H_{\text{F160W}}$ observations. The H band covers restframe emission around 4000 $\AA$, depending on the redshift of the sources and it was the detection band of the CANDELS catalogs.

\subsection{Spectroscopic catalogs}\label{sec:spec_cata}
We assembled a catalog containing all the \lya\ spectral information both in emission and absorption of galaxies at $z \geq 2$ in the GOODS-S, UDS and COSMOS fields (see Tab.~\ref{tab:fields}). In the following we briefly describe the surveys that we considered (see also Tab.~\ref{tab:survey}).\\
\begin{itemize}
    \item The ESO public spectroscopic survey VANDELS: the final data release provides redshifts and spectra for 2087 galaxies  in the CDFS and UDS fields in the range 1 < z < 6.5 \citep[][]{McLure2018, Pentericci2018, Garilli2021}. We took all data associated to spectral quality flags $QF \geq 2$, i.e. a redshift reliability $\geq$ 80\%. Line fluxes and EWs were derived using Gaussian fit measurements performed with \texttt{slinefit} \citep{Schreiber2018}. 
    The complete emission line catalogs for the VANDELS sources will be published in \cite{Talia2023}. 
    \item VUDS, the VIMOS Ultra Deep Survey \citep[][]{Cassata2015, LeFevre2015, Tasca2017} targeted star-forming galaxies at $2 \leq z \leq 6$ in the COSMOS, VVDS and CDFS fields. The \lya\ emission line were measured manually using the IRAF \texttt{splot} tool, integrating the area encompassed by the line and the continuum. We considered only galaxies with quality flags $QF \geq 3$, i.e. sources associated to a redshift reliability $\geq$ 95\%, by taking into account the line fluxes and EWs given by the team.
    \item MUSE-Wide and MUSE-Deep: these programs  targeted two different fields, COSMOS and CDFS, providing  spectroscopic data for  2052 confirmed emission line galaxies at $1.5 < z < 6.4$ \citep{Schmidt2021}. Line fluxes and EWs were extracted through Gaussian fits by the team. We used all sources  with a confidence flag greater than 1, referring to line emitters with at least a single trustworthy line detection.
    \item CANDELSz7 \citep{Pentericci2018}: this program aimed at spectroscopically confirming a homogeneous sample of $z \sim 6$ and $z \sim 7$ star forming galaxies. Candidates were selected in the GOODS-S, UDS and COSMOS fields. The \lya\ flux was measured by means of a Gaussian fit, while to determine the EW, the continuum was obtained directly from the broad band images. We included all galaxies from this sample, regardless of QF.
    \item The ESO GOODS-South follow-up: to complement the previous published spectroscopic catalogs we exploited all data  available in the ESO archive for high redshift galaxies in this field. Data were obtained by several surveys, as follow up of the GOODS-South project including GMASS \citep{Kurk2013}, GOODS-S FORS \citep[][]{Vanzella2008}, GOODS-S VIMOS-LR and GOODS-S VIMOS-MR \citep[][]{Popesso2009, Balestra2010}. For the above programs we only found published values for spectroscopic redshifts and quality flags, but no measurement of the \lya\ line. We therefore derived the \lya\ line flux and EW directly from the 1-Dimensional spectra with a Gaussian fit (see Sec.~\ref{sec:LyA_measure} for details).
    \item DEIMOS 10K: this survey \citep{Hasinger2018} targeted the COSMOS field. For each source it provides the associated spectroscopic redshift. We directly measured the \lya\ line information from the 1-Dimensional spectra with a Gaussian fit (Sec.~\ref{sec:LyA_measure}).
    \item zCOSMOS-deep: this survey targeted star-forming galaxies in the range $1.4 < z < 3.0$ \cite{Lilly2007, Kashino2022}, Lilly et al. in prep. We obtained the redshift and QF information directly from the team (private communication). We used a Gaussian fit to extract \lya\ line flux and EW directly from the 1-Dimensional spectra (Sec.~\ref{sec:LyA_measure}).
\end{itemize}
There were few cases in which the spectroscopic information of a source was reported multiple times. In this case just one estimate of EW and \lya\ flux was retained. To select  duplicated sources (i.e. the ones with the same CANDELS ID) we applied the following criteria listed in order of importance:\\
\begin{enumerate}
\item A non-detection ($EW=-99$. and flux$=-99$) was always discarded if we had any other measurement on the same galaxy. In this case only the latter was retained.
\item Sources with VANDELS quality flags 4 and 3 had the priority on all
other detection.
\item Galaxies associated with MUSE confidence flags 3 and 2 were preferred.
\item Data in CANDELSz7 catalog had then the priority.
\item The spectral information obtained through our Gaussian fit analysis was taken.
\end{enumerate}
Our final sample is composed of  1578 unique galaxies in the redshift range $z \in [2,7.9]$, whose \lya\ line EW, physical and morphological parameters are measured (Sec.~\ref{sec:method}).

\section{Methods}\label{sec:method}

\subsection{Measurements of physical properties }\label{sec:SED_fitting}
Physical properties were originally estimated by the CANDELS collaboration \citep{Santini2015}. However given the availability of many new spectroscopic redshifts obtained in the past years we re-evaluated them following the method outlined in \cite{Santini2022CANDELS}, fixing the redshift of each source to the current spectroscopic measurement when available, or to the photometric one \citep[][]{Merlin2021, Kodra2023}. We measured the stellar mass (Mass), the star formation rate (SFR), dust reddening $E(B-V)$, metallicity (Z) and Age by fitting synthetic stellar templates to the photometry of the sources with the SED fitting code \textsc{zphot} \citep{fontana00}.
We adopted \cite{Bruzual2003} models, the \cite{Chabrier2003} IMF and assumed delayed star formation histories (SFH($t$) $\propto (t^2/\tau) \cdot \exp(-t/\tau)$), with $\tau$ ranging from 100 Myr to 7 Gyr. The age could vary between 10 Myr and the age of the Universe at each galaxy redshift, while metallicity assumed values of 0.02, 0.2, 1 or 2.5 times Solar metallicity. For dust extinction, we used the \cite{Calzetti2000} law with $E(B-V)$ ranging from 0 to 1.1. Nebular emission was included following the prescriptions of \cite{Castellano2014} and \cite{Schaerer2009}. 

\subsection{ \lya\ emission line measurements}\label{sec:LyA_measure}
For the ESO GOODS-S, DEIMOS 10K and zCOSMOS-deep surveys (see Sec.~\ref{sec:spec_cata}) a measurement of the \lya\ EW is not available. We therefore measured it from the spectra. The latter were obtained from the ESO archive, from the COSMOS data access website\footnote{https://irsa.ipac.caltech.edu/data/COSMOS/spectra/deimos/} and by private communication respectively. We fitted a single Gaussian profile on the \lya\ lines in emission or in absorption using \textsc{MPFIT} \citep{MPFIT}. The code requires the 1-dimensional spectrum and the spectroscopic redshift of the source. The latter was used to get an estimate of the portion of the spectrum to fit near $\lambda_{Ly\alpha}^{obs}$: only the range  $[\lambda_{Ly\alpha}^{obs}-300, \lambda_{Ly\alpha}^{obs}+300]\AA$ was used for the fit. Furthermore 2 out of 3 free parameters of the Gaussian profile were constrained to be within the following ranges: the mean, $\mu \in [\lambda_{Ly\alpha}^{obs}-25, \lambda_{Ly\alpha}^{obs}+25]\AA$ and the standard deviation $\sigma \leq 3000$ km/s. The third parameter, i.e. the maximum flux, was left free. Note that we allow $\mu$ to vary within the defined range, because we have no information on which feature was used for the spectroscopic redshift identification. 
To consider the possible  absorption of the continuum at  wavelengths shorter than $\lambda_{Ly\alpha}^{obs}$, whenever the median of the blue continuum flux (i.e. $\lambda^{obs} < \lambda_{Ly\alpha}^{obs}$) was dimmer than the median of the red continuum by more than a standard deviation, the spectrum (including the continuum) was fitted in the range $[\lambda_{Ly\alpha}^{obs}, \lambda_{Ly\alpha}^{obs}+300] \AA$.
We want to highlight that the lines were fitted with a single Gaussian profile, with no clear cases suggesting the presence of a significant asymmetric line shape. This is also due to the medium-to-low resolution of all  spectra. For each spectrum we then visually inspected the final fit result. 
Visual inspection assures the quality of the results obtained and the lack of AGN emitters \citep[i.e. sources with strong CIV $1548 \AA$ or NV $1239 \AA$ emission lines][]{Taniguchi2005} which we might have missed to remove with the X-ray identification.

\section{The physical properties of the selected population}\label{sec:discussion}

\begin{figure}[t]
\centering
\includegraphics[trim={0.1cm 0.1cm 0.1cm 0.1cm},clip,width=\linewidth, height=6.5cm]{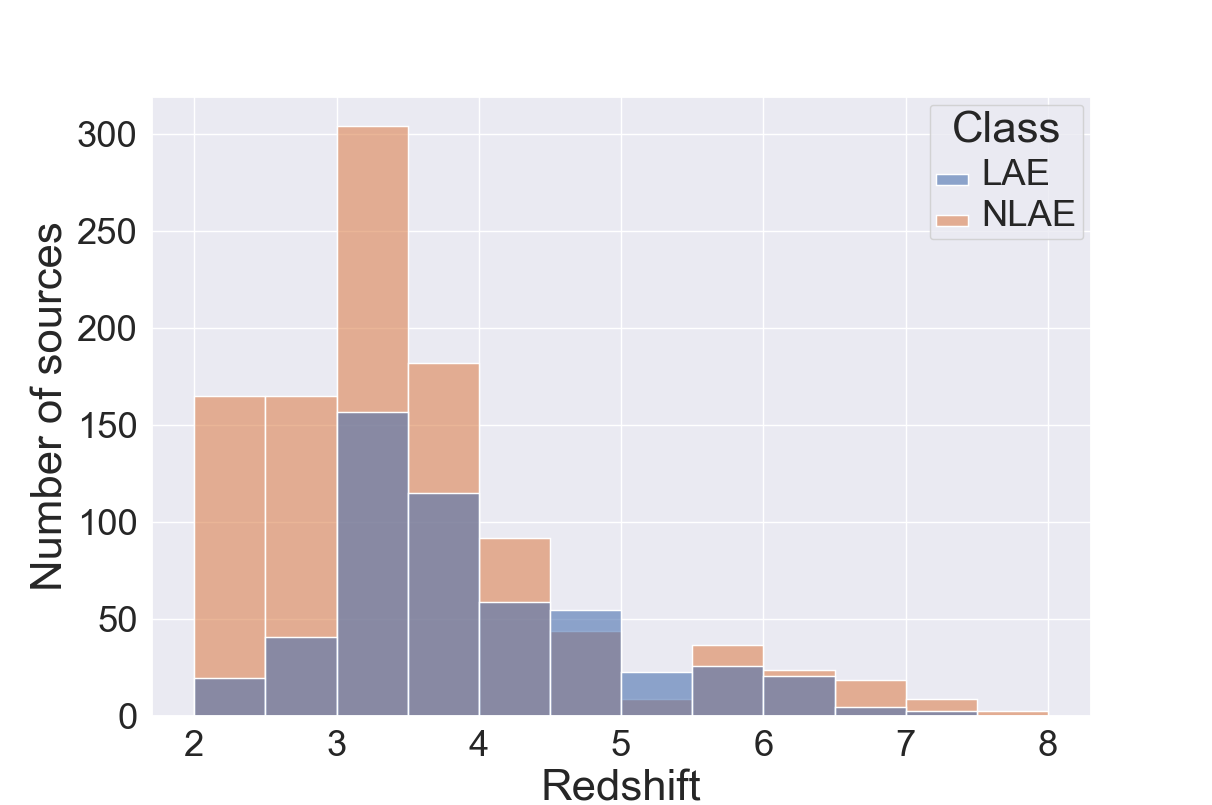}
\caption{LAEs (blue) and NLAEs (red) populations considered in the redshift range $z \in [2,7.9]$.} \label{fig:redshift}
\end{figure}

\begin{figure}[t]
\centering
\includegraphics[trim={0.1cm 0.1cm 0.1cm 0.1cm},clip,width=\linewidth, height=5.5cm]{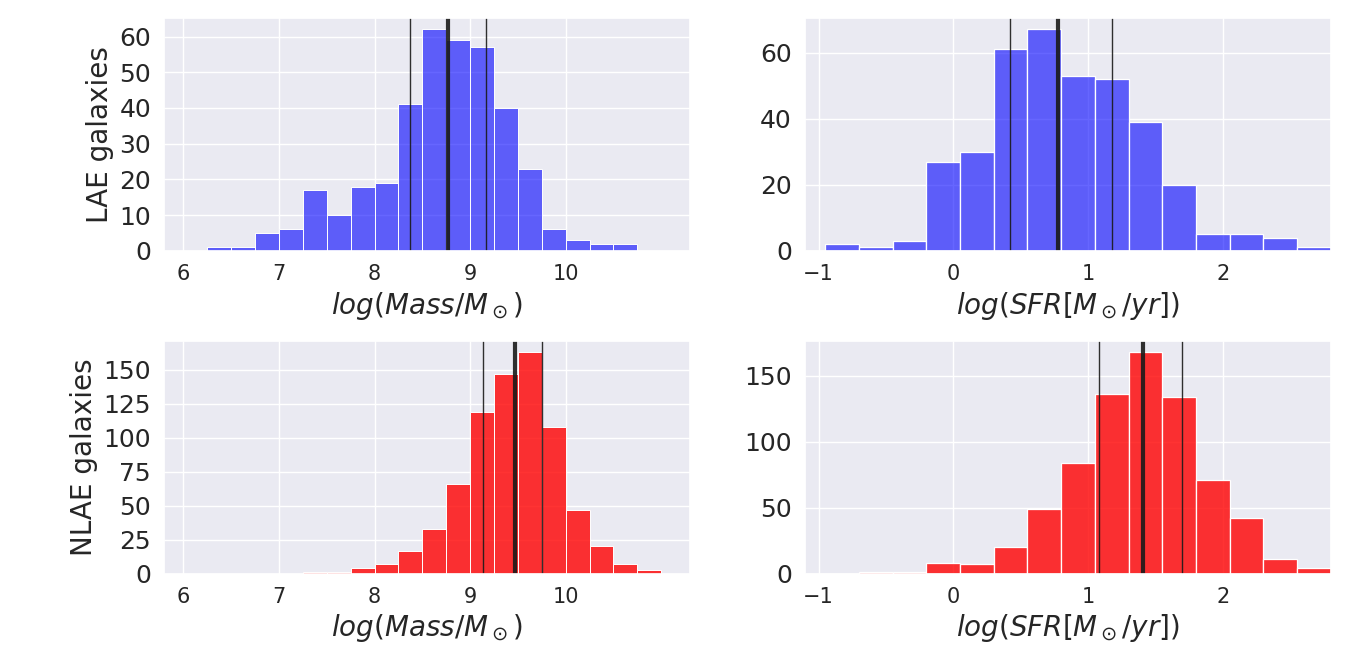}
\includegraphics[trim={0.1cm 0.1cm 0.1cm 0.1cm},clip,width=\linewidth, height=5.5cm]{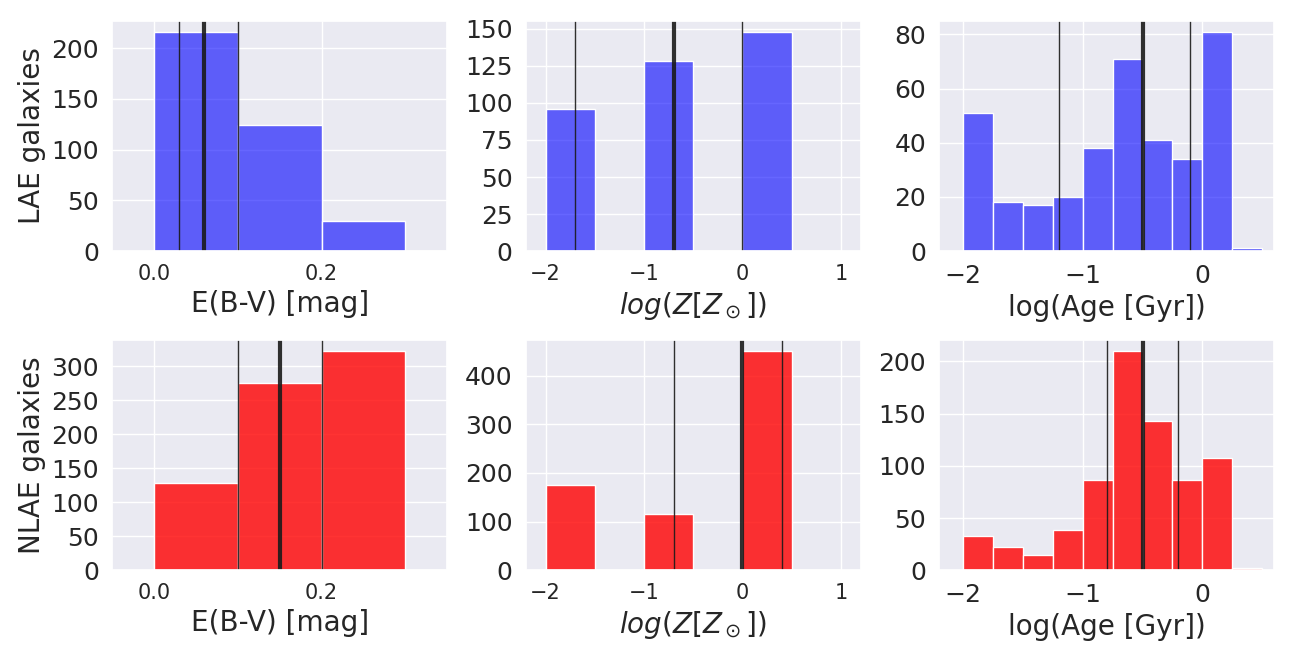}
\caption{The distributions of the physical parameters of the 1115 galaxies in the redshift range $z \in [2.5, 4.5]$. We show a direct comparison between LAEs (blue) and NLAEs (red). The thick and thin black lines correspond to the median, 25 percentile and 75 percentile values of the distributions.} \label{fig:Phys_LAEvsNLAE}
\end{figure}

\begin{figure}[t]
\centering
\includegraphics[trim={0.1cm 0.1cm 0.1cm 0.1cm},clip,width=\linewidth, height=5.5cm]{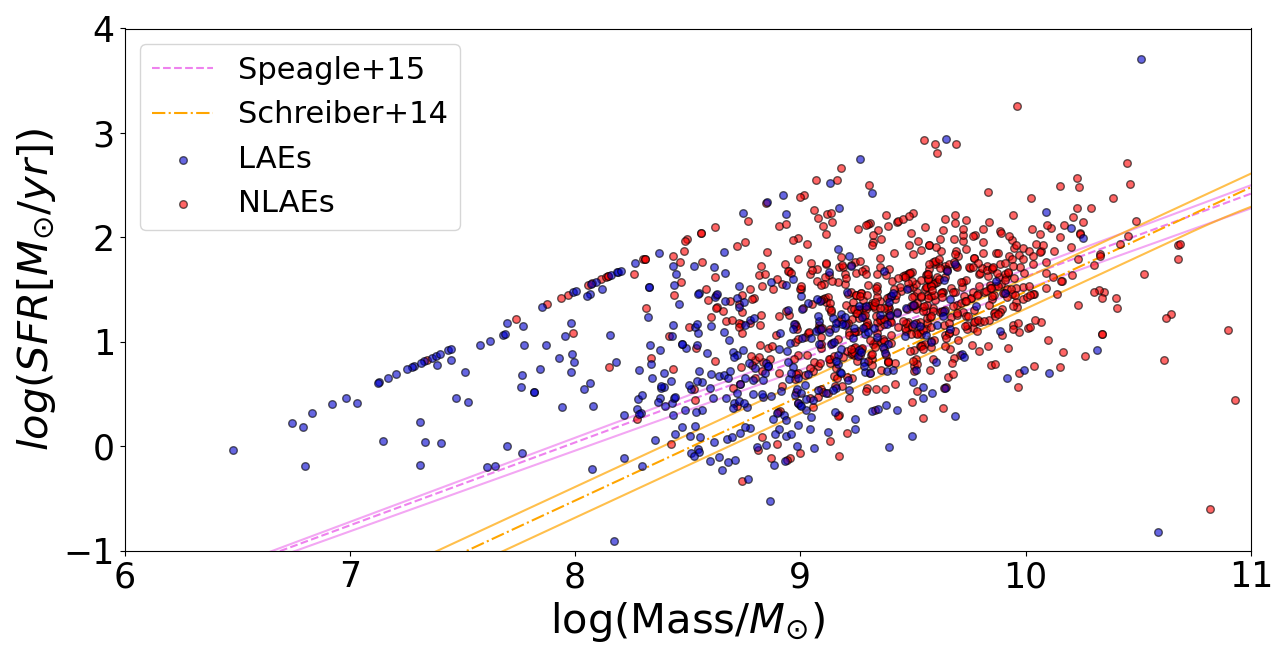}
\caption{Main Sequence diagram of the 1115 galaxies in the redshift range $z \in [2.5, 4.5]$. The yellow dot-dashed line indicates the best fit relation \cite{Schreiber2015} at $z=3.5$, the continuous yellow lines refer to the fits at $z=2.5$ and $z=4.5$. The pink dashed line indicates the fit by \cite{Speagle2014} at $z=3.5$, while the continuous pink lines refer to the fits at $z=2.5$ and $z=4.5$.} \label{fig:MainSeq_LAEvsNLAE}
\end{figure}

\begin{figure}[t]
\centering
\includegraphics[trim={0.1cm 0.1cm 0.1cm 0.1cm},clip,width=\linewidth, height=5.5cm]{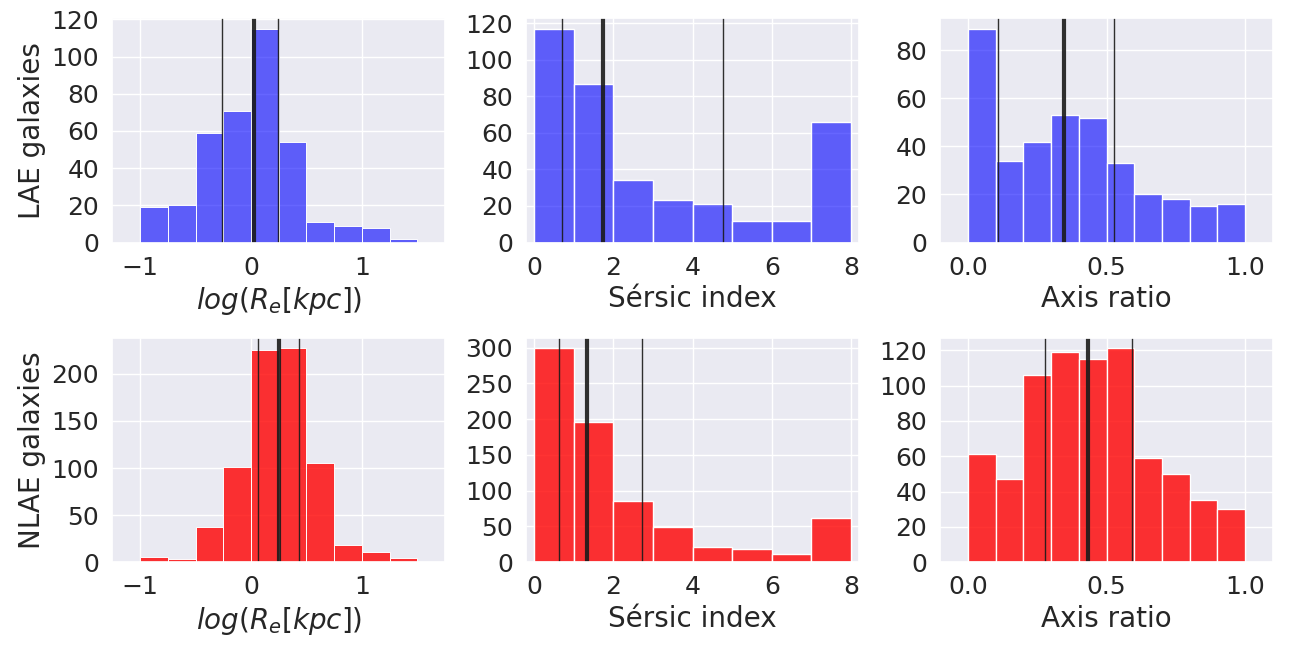}
\caption{Same as Fig.~\ref{fig:Phys_LAEvsNLAE} but for the morphological properties $R_e$, $n$ and $q$.} \label{fig:Morph_LAEvsNLAE}
\end{figure}

\begin{figure}[t]
\centering
\includegraphics[trim={0.1cm 0.1cm 0.1cm 0.1cm},clip,width=\linewidth, height=5.5cm]{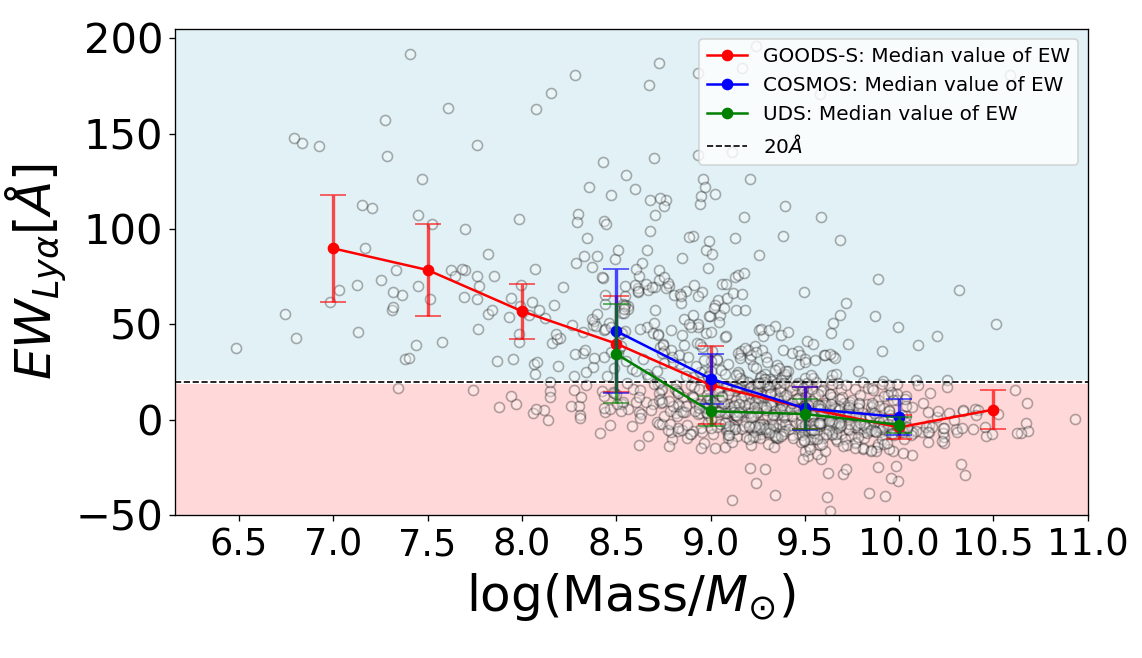}
\caption{\lya\ EW vs total stellar mass estimates. The horizontal black dashed line shows the $20\AA$ threshold a source has to exceed to be considered a LAE. For readability, the blue and red portions of the figure mark LAE and NLAE populations respectively. Red, blue and green trends represent respectively the EW median values in Mass bin of 0.5 dex for GOODS-S, COSMOS and UDS fields. They are associated with an error bar, accounting for the Median Absolute Deviation.} \label{fig:EWvsMass}
\end{figure}

\begin{figure}[t]
\centering
\includegraphics[trim={0.1cm 0.1cm 0.1cm 0.1cm},clip,width=\linewidth, height=5.5cm]{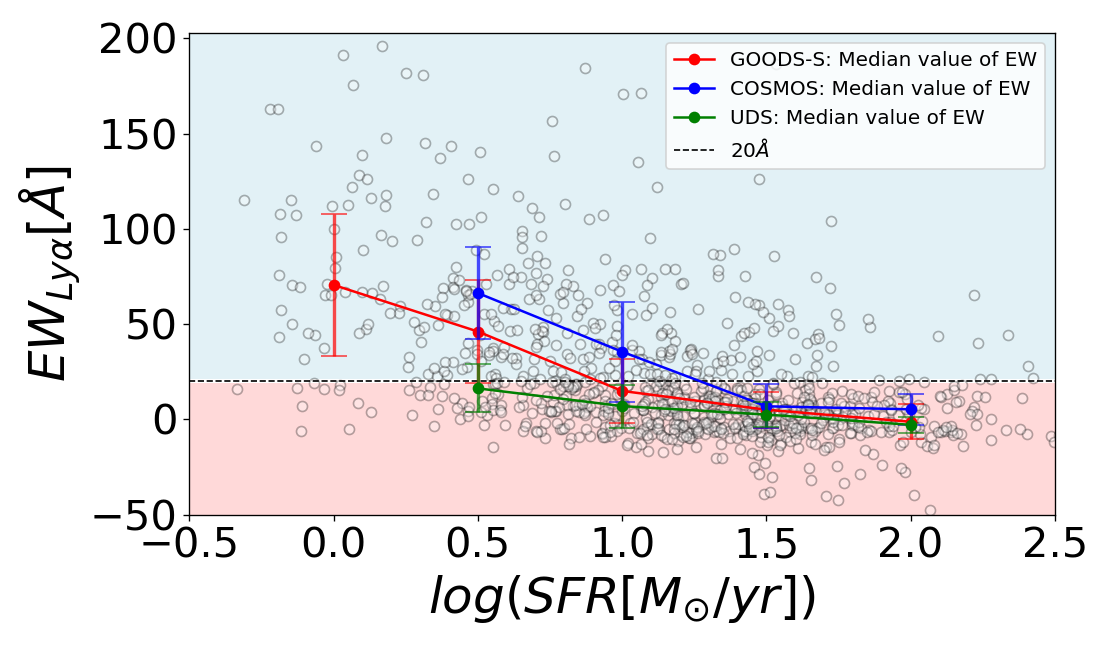}
\caption{\lya\ EW vs star formation rate. Symbols and colours are described in Fig.~\ref{fig:EWvsMass}.} \label{fig:EWvsSFR}
\end{figure}

\begin{figure}[t]
\centering
\includegraphics[trim={0.1cm 0.1cm 0.1cm 0.1cm},clip,width=\linewidth, height=5.5cm]{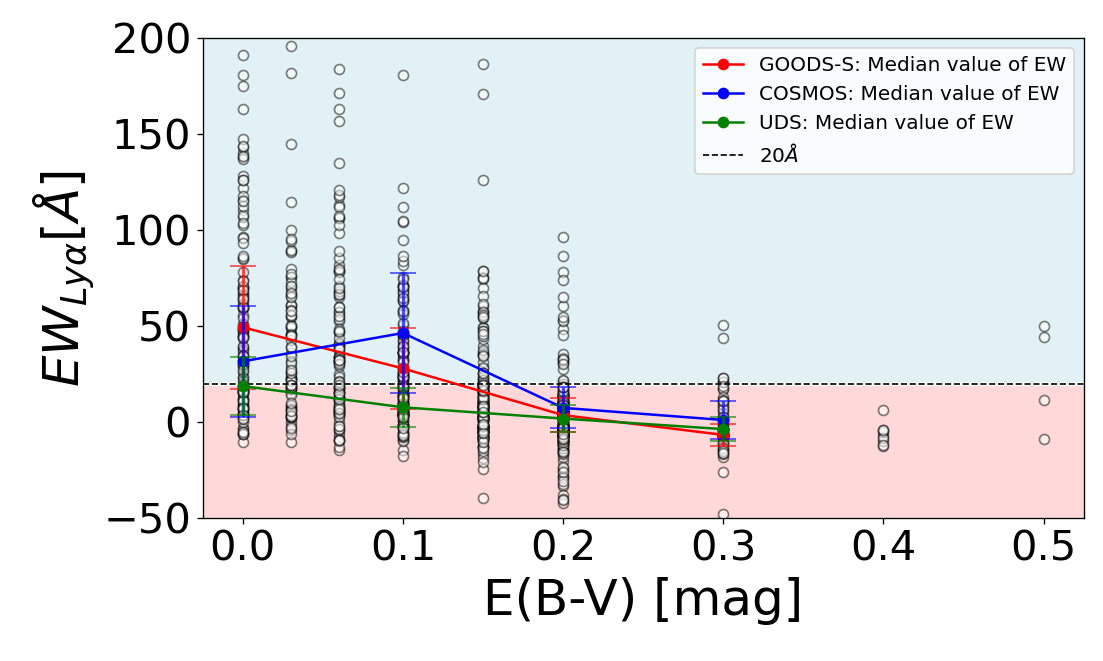}
\caption{\lya\ EW vs E(B-V). Symbols and colours are described in Fig.~\ref{fig:EWvsMass}.} \label{fig:EWvsEBV}
\end{figure}

\begin{figure}[t]
\centering
\includegraphics[trim={0.1cm 0.1cm 0.1cm 0.1cm},clip,width=\linewidth, height=5.5cm]{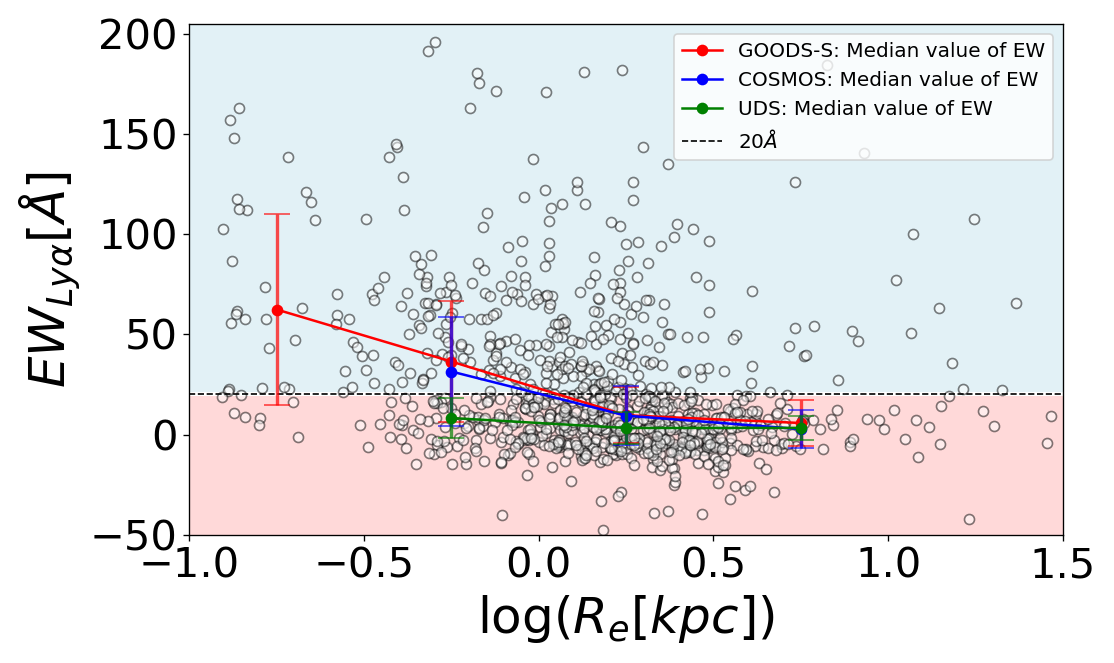}
\caption{\lya\ EW vs half-light semi-major axis $R_e$. Symbols and colours are described in Fig.~\ref{fig:EWvsMass}.} \label{fig:EWvsRe}
\end{figure}

\begin{figure}[t]
\centering
\includegraphics[trim={0.1cm 0.1cm 0.1cm 0.1cm},clip,width=\linewidth, height=5.5cm]{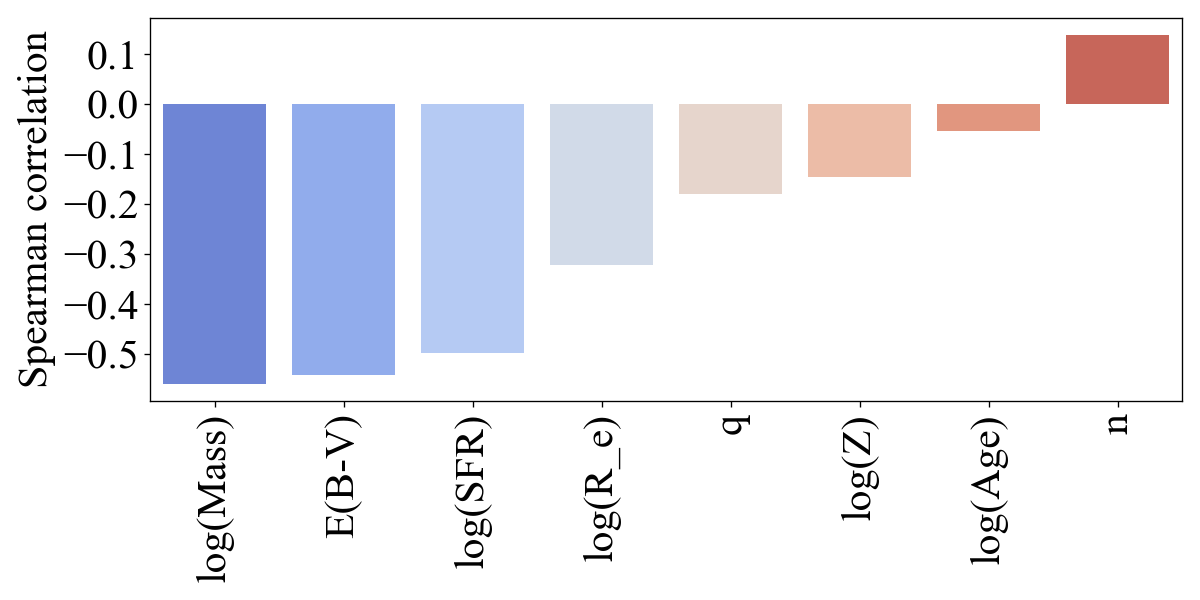}
\caption{Spearman correlation coefficients between the \lya\ EW and each physical and morphological feature, calculated on the 1115 galaxies which reside in the redshift range $z \in [2.5, 4.5]$ and are associated to a direct measure of \lya\ EW.} \label{fig:SpearmanCoeff}
\end{figure}

According to the standard definition, we consider as LAEs all the galaxies which have a rest-frame \lya\ equivalent width of $EW_0 \geq 20 \AA$ \citep[see][]{Shibuya2019, Ouchi_2020, Runnholm2020}, whilst the remaining sample is composed of NLAEs. In Tab.~\ref{tab:fields} and Tab.~\ref{tab:survey} we report respectively the total number of galaxies in the different CANDELS fields and in the spectroscopic surveys employed. Tab.~\ref{tab:survey} also presents for each survey the limiting (3 $\sigma$) flux $f_{lim}$ and redshift range targeted. In Fig.~\ref{fig:redshift} we show the redshift distribution of the whole population, with the LAEs indicated in blue and the NLAEs in red. Note that from the two distributions, it can be clearly seen that at very high redshift ($z \geq 4.0$) galaxies preferentially show \lya\ in emission. This is due both to a real effect, since as found by e.g.  \cite{Stark2010} and \cite{Cassata2015} galaxies tend to have increasingly brighter \lya\ emission as we move to earlier epochs, but also to an observational effect since it is easier to confirm the spectroscopic redshift of a galaxy if the spectrum presents a bright emission line. In our data set we can also see that the dominance of the LAEs fraction ends at $z \sim 6.5$, where it is known that the IGM is still highly neutral and effectively suppresses the \lya\ photons \citep[][]{Zheng2010, Ouchi2010, Pentericci2011, Jensen2013}. Therefore above this redshift (which is considered a proxy of the cosmic time at the end of the Epoch of Reionization) LAEs become again rarer.
\begin{table}[]
\caption{CANDELS fields selected and data sample in the redshift range $z \in [2,7.9]$. We consider LAEs as galaxies with $EW_0 \geq 20 \AA$.}\label{tab:fields}
$$
\begin{array}{lccc}
\hline \hline
\noalign{\smallskip}
    \textbf{Field} &
    \textbf{Galaxies} &
    \textbf{LAEs} &
    \textbf{NLAEs} \\ \hline
\noalign{\smallskip}
 \hline
 \noalign{\smallskip}
\textbf{GOODS-S} & 841 & 340 & 501 \\
\noalign{\smallskip}
\hline
\noalign{\smallskip}
\textbf{COSMOS}  & 408 & 107 & 301 \\
\noalign{\smallskip}
\hline
\noalign{\smallskip}
\textbf{UDS}     & 329 & 78  & 251  \\ 
\hline \hline
 \noalign{\smallskip}
 \textbf{TOTAL}                  & 1578 & 525  & 1053 \\
 \noalign{\smallskip} \hline
\end{array}
$$
\end{table}
%
%
%
\begin{table}[]
\caption{Spectroscopic surveys with 3 $\sigma$ limiting flux in units of 10$^{-18}$ erg s$^{-1}$cm$^{-2}$ and redshift range.}\label{tab:survey}
$$
\begin{array}{lcccc}
\hline \hline
\noalign{\smallskip}
    \textbf{Survey} &
    \textbf{LAEs} &
    \textbf{NLAEs} &
    \textbf{f$_{lim}$} &
    \textbf{Redshift} \\ 
\hline
\noalign{\smallskip}
\textbf{VANDELS}                & 143 & 472 & 1.2 & [2.9,6.1]\\
\noalign{\smallskip}
\hline
\noalign{\smallskip}
\textbf{VUDS}                   & 21  & 141 & 5.2 & [2.0,6.0]\\
\noalign{\smallskip}
 \hline
 \noalign{\smallskip}
 \textbf{MUSE-Wide}             & 232 & 39 & 7.8  & [3.0,6.3]\\
 \noalign{\smallskip}
 \hline
 \noalign{\smallskip}
 \textbf{MUSE-Deep}             & 23 & 8 & 0.43  & [2.9,6.4]    \\
 \noalign{\smallskip}
 \hline
 \noalign{\smallskip}
 \textbf{CANDELS-z7}            & 30  & 79  & 1.9  & [5.4,7.9]    \\
 \noalign{\smallskip}
 \hline
 \noalign{\smallskip}
 \textbf{GMASS}                 & 1   & 19  & 2.5  & [2.0,2.9]    \\
 \noalign{\smallskip}
 \hline
 \noalign{\smallskip}
 \textbf{GOODS-S VIMOS-LR}            & 12  & 71 & 6.7  & [2.5,3.0]    \\
 \noalign{\smallskip}
 \hline
 \noalign{\smallskip}
 \textbf{GOODS-S VIMOS-MR}            & 13  & 33 & 2.5  & [3.0,3.9]    \\
 \noalign{\smallskip}
 \hline
 \noalign{\smallskip}
 \textbf{GOODS-S FORS}          & 5  & 10 & 3.7  & [4.0,6.2]    \\
 \noalign{\smallskip}
 \hline
 \noalign{\smallskip}
 \textbf{DEIMOS 10K}                & 14  & 27  & 6.6 & [3.3,6.0]    \\
 \noalign{\smallskip}
 \hline
 \noalign{\smallskip}
 \textbf{zCOSMOS-Deep}          & 31  & 154 & 6.0 & [2.0,3.7]    \\ 
 \hline \hline
\end{array}
$$
\end{table}
%
We analysed the physical properties of the 1115 galaxies which reside in the redshift range $z \in [2.5, 4.5]$. 
This redshift range was chosen to be broad enough to get good statistics and, at the same time, to avoid possible considerable evolution of the intrinsic \lya\ emission properties and the effect of the increasing neutral hydrogen fraction in the IGM at $z \geq 5.5$. 
Most importantly in this redshift range our sample is 99\% complete for the identification of \lya\ emission with EW$\geq$ 20 \AA\ down to the magnitude limit of each survey (see Fig.~\ref{fig:limMag606}). The exception is the MUSE-Wide survey whose limiting \lya\ EW is higher than $20$ \AA\ for the faintest galaxies in the sample.
In the first and second rows of Fig.~\ref{fig:Phys_LAEvsNLAE}, we show the distribution for the 5 physical parameters determined in Sec.~\ref{sec:SED_fitting} for LAEs and NLAEs respectively. Histograms were designed to easily compare the two populations; in each plot the black lines correspond to the median values of the distributions. From these figures we can see that overall LAEs show smaller stellar mass, smaller SFR, lower values of $E(B-V)$ and metallicity than NLAEs. 
In Fig.~\ref{fig:MainSeq_LAEvsNLAE} we also show the stellar mass vs SFR (main sequence) relation. Both populations are mainly comprised of star forming galaxies along the main sequence, in agreement with the best fit relation found by \cite{Schreiber2015} and \cite{Speagle2014} at the same redshifts. However the LAE population tends to gather in the region described by Mass $\leq 10^9 \text{M}_{\odot}$ and SFR $\leq 10^{0.5} \text{M}_{\odot}/\text{yr}$. This is in agreement with the known properties of LAEs found in literature \citep[e.g.,][]{Fynbo2001, Nakajima2012, Hagen2014, Ouchi_2020}. Note that in the upper left region of the main sequence relation the galaxies with the highest sSFR are clustered together in a filament-like structure. This is a spurious effect, which is due to the choice of the minimum age considered (10 Myr) when using \textsc{zphot}.
A similar analysis was conducted for morphological properties whose distributions are shown in Fig.~\ref{fig:Morph_LAEvsNLAE} separately for LAEs and NLAEs: LAEs are more compact galaxies, i.e. they have smaller $R_e$, and tend to have smaller projected axis ratios than NLAEs.
\\
In Figures ~\ref{fig:EWvsMass}, ~\ref{fig:EWvsSFR}, ~\ref{fig:EWvsEBV} and ~\ref{fig:EWvsRe} we present the variation of the \lya\ EW as a function of the stellar mass, SFR, reddening and half-light semi-major axis $R_e$. 
For reference in each plot we indicate with a horizontal black dashed line the $EW=20\AA$ threshold.\\
In Fig.~\ref{fig:EWvsMass} we show that the stellar mass tends to be higher for sources with lower \lya\ EW. We also show median values in mass bins of 0.5 dex separately for GOODS-S (in red), COSMOS (in blue) and UDS (in green) to check if any systematic is present. The plot shows that GOODS-S, being the field with the deepest photometry comprises many low mass faint galaxies. LAEs are typically galaxies with stellar mass lower than $10^9 \text{M}_{\odot}$, similar to what reported  by \cite{Ouchi_2020}. We also find there are few LAEs which have stellar mass in excess of $M \geq 10^{10} \text{M}_{\odot}$, in good agreement with the results presented by \cite{Hagen2014}.\\
From Fig.~\ref{fig:EWvsSFR} we can see that LAEs typically are galaxies which show star formation rate lower than few solar masses per year $\leq 10^{0.5}-10$ $ \text{M}_{\odot}/\text{yr}$. This result is in agreement with the one reported by \cite{Ouchi_2020} ($\sim 1 - 10$ $ \text{M}_{\odot}/\text{yr}$). 
From Fig ~\ref{fig:EWvsEBV} we can see that as expected  the emission line becomes progressively fainter as the dust content increases. LAEs are sources which on average show little dust content, with typical reddening $E(B-V) \sim 0 - 0.2$ and a median value of 0.06, since \lya\ can be easily suppressed by dust present in a galaxy. A null reddening is reported by \cite{Ono2010} on a population of $\sim 600$ LAEs selected with narrow-band techniques, while \cite{Kojima2017} found the same reddening range obtained in our work. However as already found by \cite{Hagen2014}, we notice that there are also some LAEs which show larger reddening values, exceeding $0.3$. This might be due for example to a displacement  between regions from which stellar and nebular flux originate, or to a non uniform distribution of dust which could differentially suppress UV photons and not Ly$\alpha$ as first discussed by \cite{Neufeld1991}. 
Finally in Fig.~\ref{fig:EWvsRe} we show the variation of the \lya\ EW as a function of the half-light semi-major axis $R_e$. 
Even though the three median trends associated to the different fields are the most scattered relations amongst the properties studied, LAEs are very compact galaxies which on average have smaller $R_e$ than NLAEs.
This is in agreement with the result found in many previous works \citep[see also][]{Taniguchi2009, Malhotra2012, Paulino-Afonso2018}. 
\\
To quantify all the correlations described above, we have run a Spearman rank test \citep[][]{Spearman04} between the \lya\ EW and the physical and morphological properties of the galaxies. Note that this test assesses whether a monotonic relation exist between two variables, without any assumption on the form of the relation. 
The relevant p-value $p(r_s)$ is the probability of the null hypothesis of absence of any correlation. We consider a correlation to be present whenever $p(r_s)< 0.01$. The results on our sample are shown in Fig.~\ref{fig:SpearmanCoeff} and reported in Tab.~\ref{tab:Spearman}: we see that all features show anti-correlations, except for the Sèrsic index which is positively correlated and for the Age, whose p-value $> 0.01$, thus the no correlation scenario could not be discarded. 
Stellar mass, reddening, star formation rate and the the half-light semi-major axis are found to be the features that correlate  more strongly  with the observed \lya\ EW. Similar results on the strong correlation of the \lya\ EW with stellar mass, SFR and dust extinction were previously found by many works \citep[e.g.][]{Kornei2010, Pentericci2010, Oyarzun2017, Du2018, Marchi2019, McCarron2022, ChavezOrtiz2023}. \\
In the next section we try to exploit the above correlations to build a ML algorithm that can identify LAEs only on the basis of the physical and morphological properties (which can be derived by multi-wavelength photometry), without the need for costly spectroscopic observations.
%
%
%
%
\begin{table}[]
\caption{Spearman correlation coefficients with the \lya\ EW. Features are ranked by increasing p-values.}\label{tab:Spearman}
$$
\begin{array}{lccc}
\hline \hline
\noalign{\smallskip}
    \textbf{Feature} &
    \textbf{Coefficient} &
    \textbf{p-value} &
    \textbf{Null hypothesis rejected} \\ \hline
\noalign{\smallskip}
 \hline
 \noalign{\smallskip}
\textbf{Mass} & -0.561 & <10^{-5} & Yes \\ 
\noalign{\smallskip}
\hline
\noalign{\smallskip}
\textbf{E(B-V)}  & -0.551 & <10^{-5} & Yes \\ 
\noalign{\smallskip}
\hline
\noalign{\smallskip}
\textbf{SFR}     & -0.509 & <10^{-5} & Yes  \\ 
\noalign{\smallskip}
\hline
\noalign{\smallskip}
\textbf{R\_e}     & -0.313 & <10^{-5} & Yes  \\ 
\noalign{\smallskip}
\hline
\noalign{\smallskip}
\textbf{q}     & -0.201 & <10^{-5} & Yes  \\ 
\noalign{\smallskip}
\hline
\noalign{\smallskip}
\textbf{Z}     & -0.158 & <10^{-5} & Yes  \\ 
\noalign{\smallskip}
\hline
\noalign{\smallskip}
\textbf{n}     & 0.120 & 10^{-5} & Yes  \\
\noalign{\smallskip}
\hline
\noalign{\smallskip}
\textbf{Age}     & -0.024 & 0.4 & No  \\
\hline
\end{array}
$$
\end{table}
\section{Random Forest classifier}\label{sec:RandomForest}

We developed a machine learning classifier that aims at distinguishing LAEs from NLAEs, i.e. a binary class problem in which the target labels (LAEs/NLAEs) are discrete. We opted for a supervised approach because we want to use all the physical, morphological and spectroscopic data available from GOODS-S, COSMOS and UDS fields. The task of learning consists in mapping features (i.e. physical and morphological properties) to classes’ labels (i.e. LAEs/NLAEs identification thanks to spectroscopic data), based on training example pairs (i.e. features and labels shown to the classifier). To train the ML methods on the classification task of selecting LAEs, we employ only the spectroscopic labels (LAEs and NLAEs) without directly using the specific information about the \lya\ line flux and EW.

\subsection{A brief overview of the Random Forest classifier}\label{sec:RandomForestOverview}
Random Forest \citep{Breiman2001} is a publicly available \textit{scikit-learn} \citep{scikit-learn} ensemble learning classifier that combines multiple decision trees to improve classification performance. Each tree in the forest is built on a random subset of the training data. During prediction, each tree votes for the class label and the final prediction of the ensemble classifier is the majority vote of all the trees. The general idea beneath a single tree classifier consists in finding the optimal set of rules to partition the space of features to distinguish data points of different classes. A single tree-classifier works in such a way that each time a new rule is applied, the data set splits into two new branches, creating a node. Because this process is recursive, the decision graph resembles the schema of an upside-down tree. The root node at the top of a decision tree contains the entire data set. At each branch of the tree, data are divided into two child nodes subsets, based on a decision boundary: one node contains data below the decision threshold and the other one includes data above it. Geometrically speaking boundaries are hyper-surfaces axes aligned in the space of features. The splitting process repeats until a predefined stopping criteria is achieved in the leaves nodes, where all data contained are finally catalogued with just one class label, i.e. the most recurrent label in the subset associated to the terminal leaf node itself. 
The fraction of samples of the same class in all the leaf nodes is also used as the class probability output for the objects classified.
The application of a single decision tree for classifying new unlabeled data consists in following the tree's branches through a series of binary decisions until a leaf node is reached. Training a decision tree algorithm on a labeled data set means finding the optimal order of rules to minimize the number of objects not correctly classified. This is done trying to maximize purity in each node, i.e. an indicator that the considered subset contains predominantly observations from a single class. To measure purity the Gini index is commonly adopted: it estimates the probability that a randomly selected source would be incorrectly classified in the subset node if its label was drawn randomly, based on the label distribution of the same data subset. \\ Single decision trees are prone to overfitting: as the splitting process progresses thanks to the rule set by the Gini index, the error on the training set will decrease, however, at some point in its growth, the tree will cease to represent the correlations within data and will reflect the noise within the training set. The core idea of the Random Forest method is therefore to introduce random perturbations into the learning procedure of an ensemble of tree classifiers, in order to get several different models from a single learning set. This is achieved during training through the hyper-parameters (i.e. internal parameters to be set by the user) of the method: their role is to control the growth of each decision tree and to introduce random perturbation in the splitting process. The final prediction is then obtained combining the results of the whole ensemble of classifiers: the class of each object is determined by a majority vote among all the trees. 
The Random Forest classifier  also outputs the class probability associated to each object through the internal method \textit{predict\_proba}. The predicted class probabilities of an input sample are computed as the mean predicted class probabilities of the trees in the forest.\\


\subsection{Training the optimal Random Forest classifier}\label{sec:RandomForestTrain}
\begin{figure}[t]
\centering
\includegraphics[trim={0.1cm 0.1cm 0.1cm 0.1cm},clip,width=\linewidth]{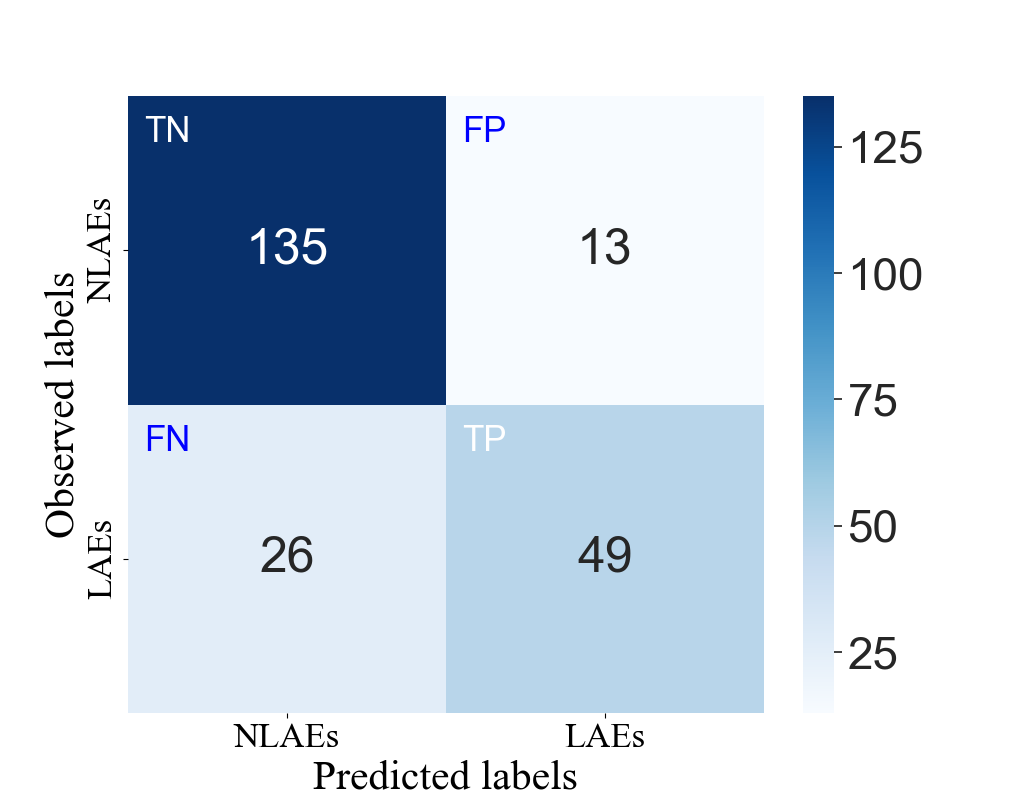}
\caption{Confusion matrix computed on the test set. From the upper left to lower right in order the total number of TN, FP, FN, TP are reported. The total number of galaxies in the test set is 223.} \label{fig:ConfMat}
\end{figure}

The search of the optimal set of hyper-parameters which fully describes the optimal Random Forest classifier was performed by a 5 \textit{k-fold cross validation}   approach and through a standard grid search. The optimal classifier is defined as the Random Forest which maximizes the cross-validation set accuracy, i.e. the average fraction of the galaxies correctly classified on the cross-validating sets. When applied to boolean data, 
using the definition of true positives (TP), false positives (FP), true negatives (TN) and false negatives (FN), the accuracy score is expressed as $\text{(TP + TN)/(TP + FP + TN + FN)}$. In our case, TP and TN are respectively the number of sources that belong to LAEs and NLAEs classes in both spectroscopic data and algorithm guess; FP is the number of sources labeled as LAEs by the algorithm but being NLAEs in the truth and finally FN is the number of sources that belong to the LAEs target class according to the spectroscopic data and miss-guessed by the ML algorithm. Precision, i.e. the fraction of truly LAEs predictions over all data predicted as LAEs by the Random Forest classifier ($\text{TP/(TP + FP)}$), is a scoring metric that was also  monitored through training and testing,  but was not considered for the search of the optimal classifier.
\\
Through the grid search we explored the most critical hyper-parameters of the method that regulate the growth of the forest during training: 
\begin{itemize}
\item \textit{n\_estimators} is the total number of trees in the forest. The more decision trees, the more opportunities the algorithm has to learn from a variety of features and subset combinations. However after exceeding a threshold, new trees do not reveal any more information because they get highly correlated with each other. This parameter varied in our grid between 50 and 600 (in multiples of fifty) 
\item \textit{max\_depth} is the maximum depth for each tree in the splitting process. This parameter, which controls the classifying capability of each tree, varied  between 5 and 30 (in multiples of five).
\item \textit{max\_features} is  the number of features to consider when looking for the best split in the splitting process. It is of key importance for growing trees slightly different from each other. Uncorrelated trees make the majority voting process of the forest more robust. Note that the forest, as an ensemble, is guaranteed to use all the features in the dataset. This parameter assumed all values between 2 and 8, i.e. up to the total number of features we have in our dataset. 
\end{itemize}
The final grid explored contains $\sim 500$ models.
A key aspect during training was setting the hyper-parameter \textit{class\_weight = "balanced\_subsample"}, such that, for every tree grown during the training, learning weights associated to the input data were inversely proportional to class frequencies. This prevents the algorithm to be biased to classify correctly only the NLAEs majority class. The other hyper-parameters, which controls the branching of each tree were left as default: \textit{min\_samples\_split} = 2, \textit{min\_samples\_leaf} = 1. The first one controls the minimum number of samples required to split a parent node into two child nodes, while the latter sets the minimum number of samples required to be at a leaf node. In other words, a split point at any depth will only be considered if the parent node holds more than \textit{min\_samples\_split} data and it leaves at least \textit{min\_samples\_leaf} training samples in each of the left and right branches. These additional requirements are also referred to as "pruning technique". By setting these two hyper-parameters to the recommended default values, we are considering an ensemble of unpruned trees (see the scikit-learn \cite{scikit-learn} website\footnote{https://scikit-learn.org/stable/modules/generated \\ /sklearn.ensemble.RandomForestClassifier.html} for a detailed reference).\\
To search for the optimal classifier the training+validating dataset has been chosen to be the $80\%$ of the 1115 galaxies in the redshift range $ z \in [2.5,4.5]$, while the remaining sources ($20\%$) have been used as a test set to check the results. The standard decision of partitioning in 5 folds the training+validating dataset assures to have a validation set's size comparable to the test data. The splitting was performed choosing a fixed random seed, such that the fraction of LAEs in the training and test set would remain similar to the percentage of LAEs ($ 33.4\%$) in the complete dataset (see Tab.~\ref{tab:train_test}). We used the same internal random seed to get reproducible results. 
The best performances achieved by maximizing the cross-validation set accuracy were obtained when setting the following hyper-parameters: \textit{n\_estimators = 500}, \textit{max\_depth = 20} and \textit{max\_features = 3}. This optimal Random Forest classifier achieves accuracy values of $(79.4 \pm 3.6)\%$ and $82.5\%$ for the cross-validation and test set respectively. 
The scores for the precision are $(74.4 \pm 8.2)\%$ and $79.0\%$ for the cross-validation and test set respectively. The uncertainty on cross-validation results is the standard deviation on the 5 folds, while the test results do not have an uncertainty, being a single sample. Maximizing precision instead of accuracy over the cross-validation set does not change significantly the results. Note that the final accuracies obtained are comparable to those reached by narrow band selected samples, whose contamination rates range from few to 30\%, depending on redshift and magnitude limit \citep{ouchi2018}. 
In Fig.~\ref{fig:ConfMat} from the upper left to lower right we report the confusion matrix values accounting for the TN, FP, FN, TP in the test set of 223 galaxies. 
In Fig.~\ref{fig:FeatImp} we highlight the most important features used for the classification task during training: the order of importance is in very good agreement with the absolute values of the Spearman correlation coefficients found between the \lya\ EW and the features analysed (Fig.~\ref{fig:SpearmanCoeff}). 
This proves that the Random Forest method builds on these correlations and succeeds in recognizing that \lya\ emitters tend to be low mass, low star forming galaxies which have little dust content and very compact sizes. The galaxies' orientation with respect to the line of sight, accounted by the projected axis ratio q, represents the middle point in order of importance of the presented correlations found both from the Spearman test within data and the Random Forest classifier. The remaining features (Sérsic index, age and metallicity) fall behind.

\subsection{Misclassified objects}\label{sec:misclassified}
\begin{figure}[t]
\centering
\includegraphics[trim={0.1cm 0.1cm 0.1cm 0.1cm},clip,width=\linewidth, height=5.5cm]{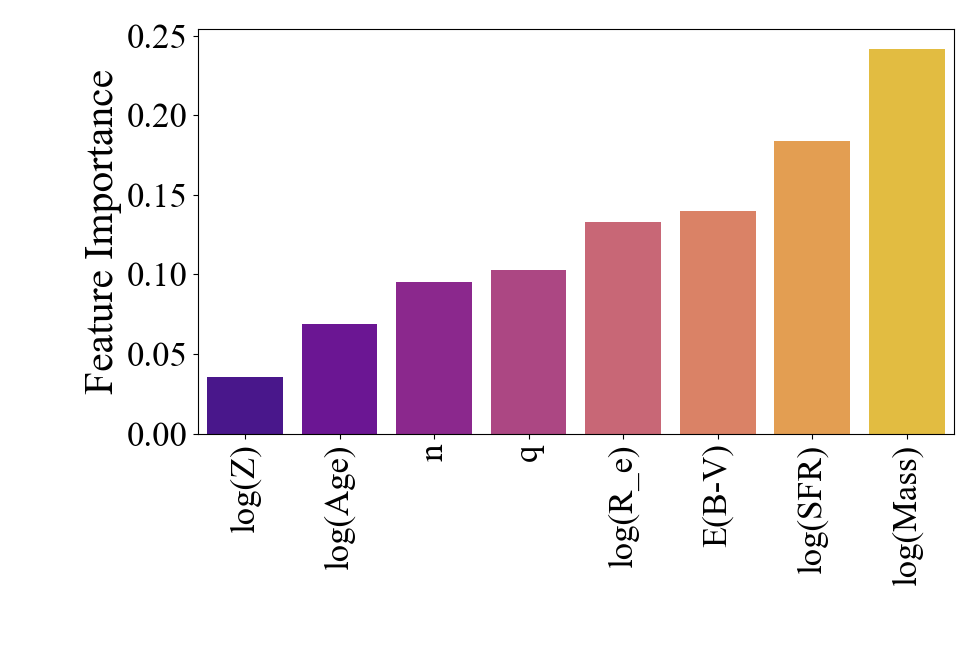}
\caption{Ranking of the most important features used by the optimal Random Forest classifier method during its training.} \label{fig:FeatImp}
\end{figure}

We further investigated the causes which make the Random Forest misclassify 26 LAEs (FN sample) by comparing the median values of this sample with the ones related to the LAEs showed during training. The FN sample and the LAEs in the training set respectively have median log(Mass) of 9.1 and 8.8, log(SFR) of 1.1 and 0.8, E(B-V) of 0.10 and 0.06, log($R_e$) of 0.08 and 0.05. The FN sample is thus composed of LAEs, whose stellar mass, SFR, reddening and half-light semi-major axis are all higher than the ones of the LAEs in the training. Since  those are the most important features (Fig.\ref{fig:FeatImp}) that the Random Forest tool is using for classification, the method gets misled and fails to recognize them. \\
A similar analysis was conducted for the 13 galaxies misclassified as LAEs (FP sample) by inspecting the most important features. These galaxies have median log(Mass) of 8.7, log(SFR) of 0.8, E(B-V) of 0.1 and log($R_e$) of 0.07, at variance with the median values of the NLAEs population shown during training (median log(Mass) of 9.4, log(SFR) of 1.4, E(B-V) of 0.15 and log($R_e$) of 0.25). The FP sample seems to be composed of peculiar NLAEs with smaller values of stellar mass, SFR, reddening and half-light semi-major axis than the ones shown during training.\\
Random Forest tends to misclassify galaxies with intermediate  properties in the test sample. To improve the results from this method, a larger  training set set would be needed. Our results would also benefit from adding further properties that could be derived from photometry and which should correlate (or anticorrelate) with the \lya\ equivalent width. For example we could add the $\xi_{ion}$ i.e. the ionizing photon production efficiency, which is given as output by some SED fitting codes \citep[e.g. BEAGLE][]{Chevallard2016} and has been found to strongly correlate with the Ly$\alpha$ emission strength, as recently shown by \cite{Castellano2023xion}. This would help recognizing LAEs from NLAEs when the other features have intermediate values, as in the cases discussed for the FP and FN samples.\\
We investigated which is the minimum probability to set when looking at all the positive predictions, in order to get only TP i.e. maximising  the purity of the predictions to 100\%. To this aim we used the \textit{predict\_proba} method provided for the classifier and described in  \ref{sec:RandomForestOverview}. We found that with a 0.93 cut in probability of being a LAE, the classifier finds only true positive (LAEs). However in this way we would lose as many as $\sim 84\%$ of the real LAEs as well, thus resulting in a very limited final sample.


\begin{table}[]
\caption{Training and Test sets considered in the search of the optimal model's hyper-parameters. }\label{tab:train_test}
$$
\begin{array}{lcccc}
\hline \hline
\noalign{\smallskip}
    \textbf{} &
    \textbf{LAEs} &
    \textbf{NLAEs} &
    \textbf{LAEs fraction} &
    \textbf{Total}
    \\ \hline
\noalign{\smallskip}
 \hline
 \noalign{\smallskip}
 \textbf{Complete data}                & 372 & 743 & 33.4\% & 1115 \\
\noalign{\smallskip}
\hline
\noalign{\smallskip}
\textbf{Train set}                & 297 & 595 & 33.3\% & 892 \\
\noalign{\smallskip}
\hline
\noalign{\smallskip}
\textbf{Test set}                   & 75 & 148  & 33.6\% & 223 \\ \hline
\end{array}
$$
\end{table}

\subsection{Testing the solidity of the optimal Random Forest}\label{sec:MLresults}
The test set result obtained by searching for the optimal Random Forest classifier is linked to the particular sub-division in the training and test datasets, performed when splitting data through the fixed random seed chosen. We thus evaluated the solidity of the optimal Random Forest (\textit{n\_estimators = 500}, \textit{max\_depth = 20} and \textit{max\_features = 3}; see Sec.~\ref{sec:RandomForestTrain}) by creating 100 different training and test sets (with $80\%$ and $20\%$ ratios) through 100 unique shuffling seeds applied to our sample of 1115 galaxies. In this iteration process we required the two subsets to have roughly the same LAEs percentage as the whole dataset. \textrm{Since we have already determined the optimal Random Forest, in this procedure we do not need a validation set}. The average results on the 100 test sets for accuracy and precision are respectively $(79.7 \pm 2.1)\%$ and $(73.1 \pm 4.3)\%$ in good agreement with the results obtained during the cross-validation process. The uncertainties reported are the standard deviations derived from the 100 test sets.

One of the purposes of training the optimal Random Forest classifier is to develop a robust method to select galaxies which have the highest probability of being \lya\ emitters from photometric catalogs. We therefore tested the optimal classifier in the case where only the photometric information is available, i.e. without the spectroscopic redshift information. For this purpose we re-evaluated the physical properties of our sample through  SED fitting by fixing the redshift of each source to the photometric estimate \citep{Kodra2023} and following the same procedure described in Sec.~\ref{sec:SED_fitting}. We then defined the training+validating (test) dataset as the 80\% (20\%) of the 1081 galaxies in the redshift range $z_{phot} \in [2.5, 4.5]$. Note that by using the photometric redshifts, we lose 34 previously considered galaxies, i.e. $\sim 3\%$ of the total. Using the same procedure described in Sec.~\ref{sec:RandomForestTrain}, we re-trained and tested the optimal Random Forest 
on this new dataset achieving a test accuracy and precision of 82.0\% and 81.1\% respectively. These values are in very good agreement with the results previously obtained both in Sec.~\ref{sec:RandomForestTrain} and Sec.~\ref{sec:MLresults}. In turn this shows that the classification does not suffer from the uncertainties derived from an SED fitting based on the photometric redshifts instead of the spectroscopic ones as also shown by several works \citep{Merlin2021, Kodra2023, ArrabalHaro2023}. 


In Sec.~\ref{sec:spec_cata} we described how we assembled the largest possible spectroscopic sample by considering 11 different observational programs. In each of these surveys targets were pre-selected using different criteria (colour selection and/or photometric redshifts). Clearly this could cause selection biases in the final sample, which are difficult to assess. To evaluate how the different selection functions could  affect our results we carried out  two different tests:
\begin{enumerate}
\item we trained and tested the optimal classifier on just one survey, therefore on a subset with a unique selection function. We considered the  VANDELS  survey  because it is the only subset with enough data (555 galaxies in the redshift range $z \in [2.5, 4.5]$) which could be then split into the training and test samples (80\% and 20\% respectively). After applying the same procedure as Sec.~\ref{sec:RandomForestTrain}, we obtained a $79.3 \%$ test accuracy, consistent with results obtained with the entire sample.
\item we trained the optimal classifier on all the data, except for a survey to be left as the test set. Note that this resembles the possible case in which the optimal classifier trained in Sec.~\ref{sec:RandomForestTrain} would be applied to an independent data set from a new survey with a different pre-selection.  We used the combined  GOODS-S VIMOS and GOODS-S FORS samples as the test set (given that they were selected in the same way). In this case we used 982 galaxies for the training and 133 for testing the performances, with a training-test ratio which is roughly 90\%-10\%. We obtained a 82.7\% test accuracy again  consistent with the results on the entire sample.  
\end{enumerate}
In conclusion, although our tests are not exhaustive, we find that the different selection effects do not change substantially the efficiency of the optimal algorithm. This is probably due to the fact that current photometric redshift codes in general are rather robust and in reasonable agreement with each other, especially for star forming galaxies in the redshift range where we carry out our training.

\subsection{Application of the optimal method to higher redshift}\label{sec:application}
Finally we applied the optimal classifier on the 194 galaxies in our initial sample with redshifts between 4.5 and 6. This sample contains a higher fraction (53.6$\%$) of LAEs compared to the one used in our training and test. As already mentioned in the introduction, this is due both to an evolution in the intrinsic galaxy properties and to  an observational effect, since spectroscopic surveys can constrain only emission lines brighter than a limiting flux from galaxies. Since galaxies get fainter at increasingly high redshift, at some point we are biased to confirm the redshift more easily in the presence of a \lya\ emission. In Fig.~\ref{fig:limMag814} we show the completeness of our sample, finding that many surveys are not complete at the faint end. 
Applying the optimal algorithm to this completely independent dataset leads to a 73.2 $\%$ accuracy and 80.2 $\%$ precision.
The lower accuracy obtained can be caused by the partial completeness in terms of spectroscopic identification of this sample.\\
In any case the high precision achieved indicates the possibility of using our trained method for identifying LAE candidates during the Reionization Epoch, where observations of large samples would be needed, e.g. to map the spatial inhomogeneous distribution of the neutral hydrogen fraction in the IGM \citep{takehro22}. Assuming a 50\% (25\%) of \lya\ transmission due to a moderately neutral (highly neutral) IGM at z$\simeq$7, if we optimally selected our spectroscopic candidates using our method which has an 80\% precision, we would obtain $\simeq$40\% (20\%) LAEs detection rate, i.e. much higher than current detection rates at z$\simeq 7$ galaxies \citep{Pentericci2018}. This would open the possibility to distinguish more easily amongst regions with high/low neutral hydrogen content in the IGM. In addition  also  other types of Ly$\alpha$ diagnostics that probe cosmic reionization,  such as emission line shape analysis \citep{Ouchi_2020} could be carried out  with much larger samples.

\section{Summary and Conclusions}\label{sec:conclusion}
Searching for LAEs at high-redshift is a challenging task due to both the limitations of the narrow-band deep imaging surveys and to the time constraints to be faced when planning a blind spectroscopic survey. In this work we presented a new efficient method based on Machine Learning which builds on the correlations found in the high redshift star forming galaxies  between the strength of the \lya\ emission and physical and morphological properties, which are easily determined from multi-wavelength photometry. The purpose is to select galaxies which have the highest probability of being \lya\ emitters. 

We initially assembled a very large sample of 1578 galaxies at $z \in [2,7.9]$ selected from the CANDELS GOODS-S, COSMOS and UDS fields, for which we have deep spectroscopic observations, including the \lya\ emission, as well as accurate physical and morphological properties derived in an homogeneous way from multi-wavelength photometry. We then considered galaxies in the redshift range $z \in [2.5,4.5]$ where statistics is higher and where  the spectroscopic surveys are complete for the identification of \lya\ emission with EW$\geq 20$\AA\ . This selected sample of 1115 sources,  is mainly formed by Main Sequence (MS) star forming galaxies, in agreement with the best fit MS relation found in literature \citep{Speagle2014, Schreiber2015}.
We find that the strength of the \lya\ emission is strongly correlated with stellar mass, dust content, SFR and half-light radius, in the sense that the line tends to be brighter for galaxies with small stellar mass, low SFR, low dust content and small radius, as already found by previous authors \citep{Taniguchi2009, Ono2010, Malhotra2012, Hagen2014, Kojima2017, Paulino-Afonso2018, Ouchi_2020}. In turn this can be explained by the major importance of the neutral hydrogen column density and the dust content within the inter-stellar medium in determining the rate of escape of \lya\ photons from a galaxy. We find that the galaxy orientation with respect to the line of sight is only mildly correlated to the \lya\ emission line. This suggests a scenario in which the preferential channels in the inter-stellar medium through which \lya\ photons escape without getting absorbed by dust are mildly dependent of the particular galaxy orientation. 

We then trained a Random Forest classifier on the task of identifying LAEs by using all the physical and morphological information available, i.e. 8 features in total. The search of the optimal Random Forest classifier was performed by a 5 \textit{k-fold cross validation} approach and through a standard grid-search. Our best results were obtained by setting the following  hyper-parameters: \textit{n\_estimators = 500}, \textit{max\_depth = 20}, \textit{max\_features = 3}, \textit{min\_samples\_split} = 2 and \textit{min\_samples\_leaf} = 1. This optimally trained classifier when applied to an independent set of galaxies in the same redshift range as the training set, recovers true LAEs with a $(79.7 \pm 2.1)\%$ accuracy and $(73.1 \pm 4.3)\%$ precision. 

The method could be further refined both by enlarging the training set to contain more numerous and more diverse galaxies, and by adding other predictive features, i.e. properties that also correlate with the \lya\ strength. One possibility could be the ionizing photon production efficiency $\xi_{ion}$, which was found to correlate with the \lya\ equivalent width \citep[][]{Harikane2018, Castellano2023xion}. This could help the method to be more robust to false classification of galaxies with intermediate properties, as also suggested by our analysis of the mis-classified objects.

When applying the classifier to a higher redshift $z \in [4.5, 6]$ dataset of 194 galaxies we obtained a slightly lower accuracy of $73.2\%$ but a precision as high as $80.2\%$. The Random Forest classifier is therefore successful at selecting LAEs at high redshift and could be used to optimally plan spectroscopic follow up observations in fields which boast good multi-wavelength photometric observations. This would allow us to maximise our chances to detect galaxies with \lya\ emission that are one of the best tools to study the epoch of reionization.
As an example our algorithm could be applied to  future  high redshift target selection with MOONS, the next generation spectrograph for the VLT which, with its large field of view  and very high multiplexing capabilities, will offer us the possibility to obtain spectra  of hundreds of high redshift galaxies in the epoch of reionization \citep{Maiolino2020}. MOONS will be able to observe the \lya\ emission throughout all phases of reionization. With a survey tailored at maximising the high redshift galaxies (z$\geq$6) which should intrinsically have strong \lya\ emission, as selected by our algorithm, we could easily distinguish between regions that have a large IGM transmission, i.e. regions that are already highly ionized, from regions where the final escaping \lya\ is very reduced due the high fraction of IGM neutral hydrogen content. We could therefore  directly analyse  the patchy spatial distribution of neutral hydrogen and compare it to the prediction of simulations.

\begin{acknowledgements}
We thank the referee for the constructive feedback provided. We acknowledge support from INAF mainstream program VANDELS. L.N. thanks Stefano Giagu and Viviana Acquaviva for giving useful insights into the Machine Learning classification.
\end{acknowledgements}

\bibliographystyle{aa}
\bibliography{biblio.bib}

\appendix
\section{Completeness of the surveys considered}\label{sec:app_completeness}
\begin{figure}[htp]
\centering
\includegraphics[trim={0.1cm 0.1cm 0.1cm 0.1cm},clip,width=\linewidth, height=5.5cm]{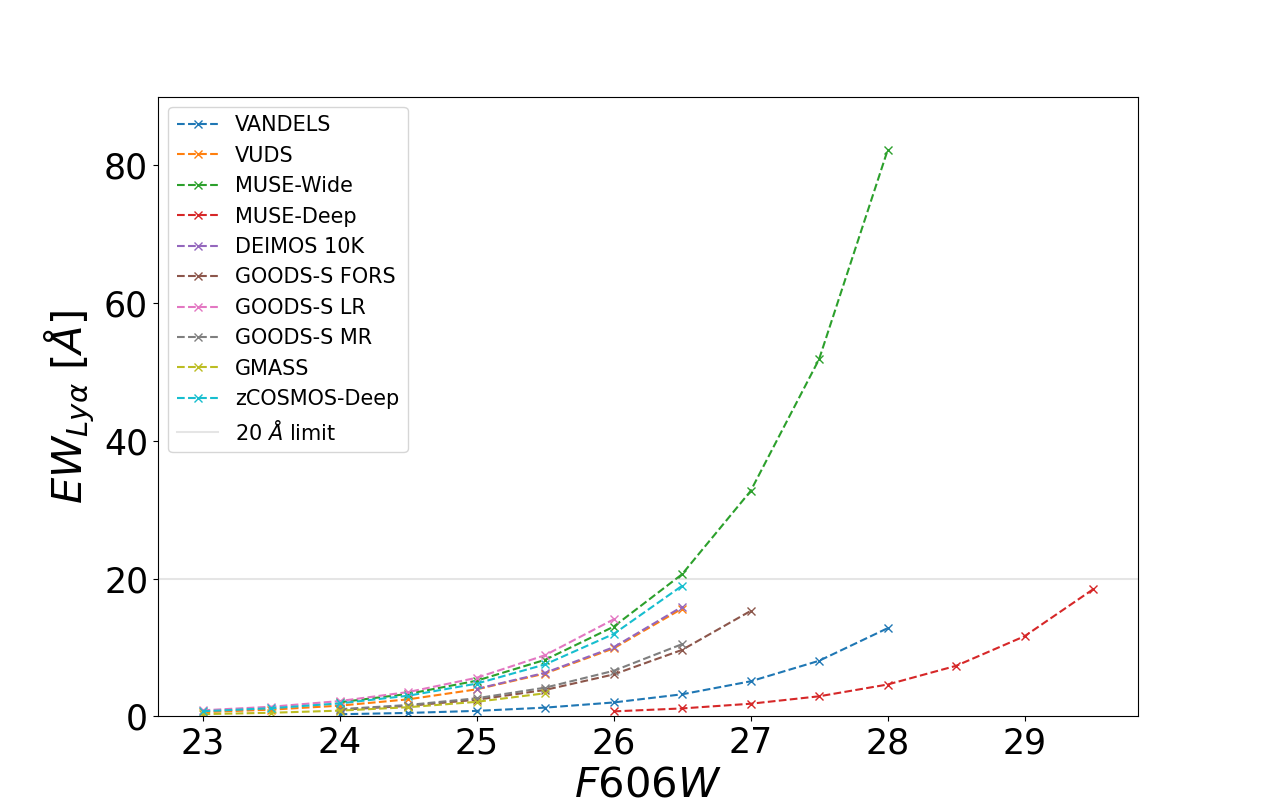}
\caption{Limiting \lya\ EW vs F606W magnitudes for galaxies at $2.5 \leq z \leq 4.5$ derived from the limiting 3$\sigma$ fluxes reported in Tab.~\ref{tab:survey}. Each survey covers a different magnitude range, according to the galaxies targeted. The grey horizontal line shows the $20 \AA$ threshold.} \label{fig:limMag606}
\end{figure}

\begin{figure}[htp]
\centering
\includegraphics[trim={0.1cm 0.1cm 0.1cm 0.1cm},clip,width=\linewidth, height=5.5cm]{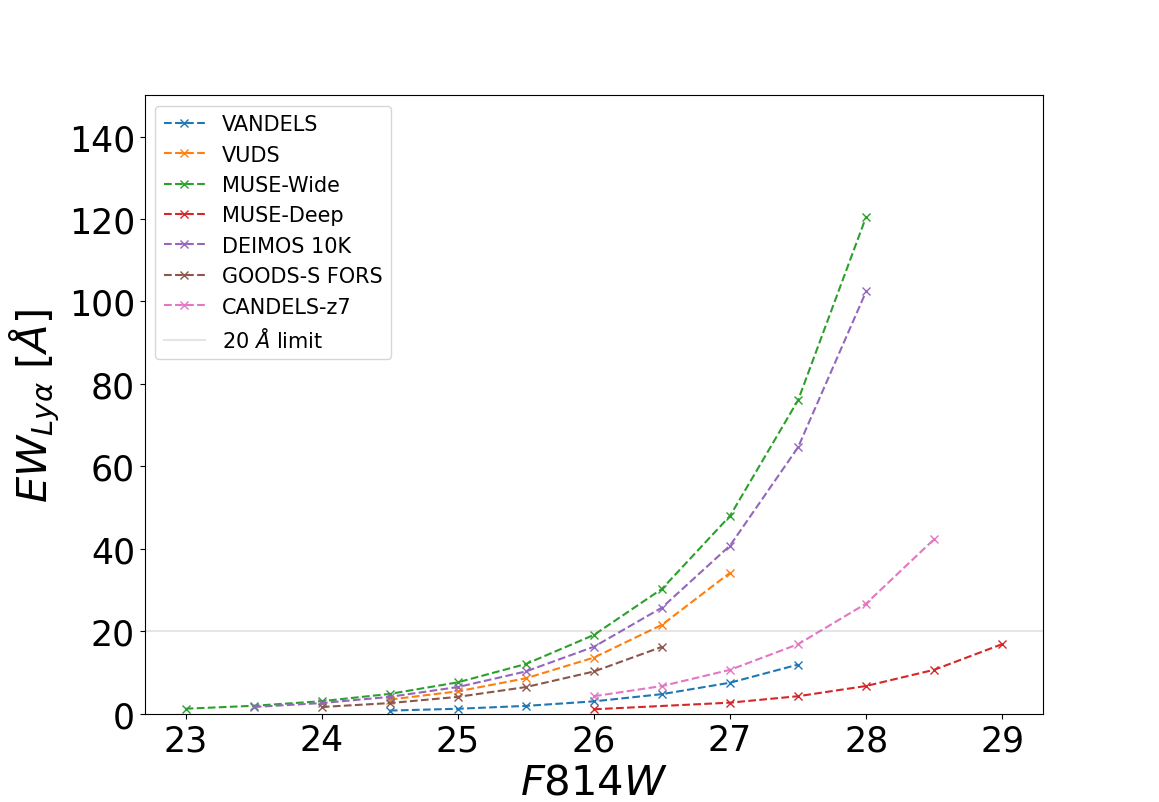}
\caption{Limiting \lya\ EW vs F814W magnitudes for galaxies at $4.5 \leq z \leq 6$. Symbols are the same as in Fig.~\ref{fig:limMag606}} \label{fig:limMag814}
\end{figure}

In this appendix we show the test performed in order to assess the completeness of our sample for each survey considered. \\
We first considered galaxies at $2.5 \leq z \leq 4.5$, i.e. the sample used for training our method. From the photometric catalogs \citep[]{Galametz2013, Nayyeri2017, Merlin2021} we associate to each survey the magnitude range spanned for the F160W HST filter, where the continuum emission near to the \lya\ line is redshifted. From the limiting (3 $\sigma$) flux $f_{lim}$ reported in Tab.~\ref{tab:survey} and assuming the mean redshift of the survey for each subsample, we then compute the limiting \lya\ EW restframe. 
In Fig.~\ref{fig:limMag606} we compare the limiting \lya\ EW with the F606W magnitude range. Overall, in this redshift range our sample is >99\% complete down to the magnitude limit of each survey. The exception is MUSE-Wide whose limiting \lya\ EW exceeds the $20 \AA$ for magnitudes $>$ 26.5. However, the number of these sources is limited compared to the 1115 galaxies in the data sample at this redshift range. Thus, losing some faint NLAEs for this survey does not affect our analysis.\\
In Fig.~\ref{fig:limMag814} we report the limiting \lya\ EW with the F814W magnitude range, for the subset of galaxies at $4.5 \leq z \leq 6$. In this case the F814W HST filter holds the information on the continuum emission near to the \lya\ line. As a result, given the limiting (3 $\sigma$) flux $f_{lim}$ reported in Tab.~\ref{tab:survey}, many surveys are not complete in the faint end population of galaxies. We decided not to include galaxies at this redshift range for training our method, because of the partial completeness.

\end{document}